\documentclass{article}

\begin{document}

\centerline{\Large \bf Performance of Equal Phase-Shift}
\centerline{\Large
\bf Search for One Iteration}



\footnote{%
The paper was supported by NSFC(Grants No. 60433050 and 60673034), the basic
research fund of Tsinghua university NO: JC2003043.}

\centerline{Dafa Li$^{a}$\footnote{email address:dli@math.tsinghua.edu.cn},
Jianping Chen$^a$, Xiangrong Li$^{b}$, Hongtao Huang$^{c}$, Xinxin Li$^{d}$ }

\centerline{$^a$ Dept of mathematical sciences, Tsinghua University, Beijing
100084 CHINA}

\centerline{$^b$ Department of Mathematics, University of California, Irvine, CA
92697-3875, USA}

\centerline{$^c$ Electrical Engineering and Computer Science Department} %
\centerline{ University of Michigan, Ann Arbor, MI 48109, USA}

\centerline{$^d$ Dept. of computer science, Wayne State University, Detroit, MI 48202,
USA}

Abstract

Grover presented the phase-shift search by replacing the selective
inversions by\ selective phase shifts of $\pi /3$. In this paper, we
investigate the phase-shift search with general equal phase shifts. We show
that for small uncertainties, the failure probability of the Phase-$\pi /3$
search is smaller than the general phase-shift search and for large
uncertainties, the success probability of the large phase-shift search is
larger than the Phase-$\pi /3$ search. Therefore, the large phase-shift
search is suitable for large-size of databases.

PACS number: 03.67.Lx

Keywords: Amplitude amplification, the \ phase-shift search, quantum
computing.


\section{Introduction}

Grover's quantum search algorithm is used to find a target state in an
unsorted database of size $N$\cite{Grover98}\cite{Grover05}. Grover's
quantum search algorithm can be considered as a rotation of the state
vectors in two-dimensional Hilbert space generated by the initial ($s$) and
target ($t$) vectors\cite{Grover98}. The amplitude of the desired state
increases monotonically towards its maximum and decreases monotonically\
after reaching the maximum \cite{LDF05}. As mentioned in \cite{Grover05}
\cite{Brassard97}, unless we stop when it is right at the target state, it
will drift away. A new search algorithm was presented in \cite{Grover05} to
avoid drifting away from the target state. Grover proposed the new algorithm
by replacing the selective inversions by\ selective phase shifts of $\pi /3$%
, the algorithm converges to the target state irrespective of the number of
iterations. In his paper, Grover demonstrated the power of his algorithm by
calculating its success probability when only a single query into the
database was allowed. It turned out that if the success probability for a
random item in the database was $1-\epsilon $, where $\epsilon $ is known to
randomly lie somewhere in the range $\left( 0,\epsilon _{0}\right) $, after
a single quantum query into the database, Grover's new \ phase-shift
algorithm was able to increase the success probability to $1-\epsilon
_{0}^{3}.$ This was shown to be superior to existing algorithms and later
shown to be optimal\cite{Grover06}\cite{Tulsi}.

In \cite{Farhi}\cite{Roland}, adiabatic quantum computation provides an
alternative scheme for amplitude amplification\ that also does not drift
away from the solution. In \cite{Tulsi}, an algorithm for obtaining fixed
points in iterative quantum transformations was presented and the average
number of oracle queries for the fixed -point search algorithm was
discussed. In \cite{Boyer}, Boyer et al. described an algorithm that
succeeds with probability approaching to 1. In \cite{LDF06}, \ we discussed
the \ phase-shift search algorithm with different phase shifts.

As discussed below, the implementation of the general \ phase-shift search\
relies on selective phase shifts. In this paper, we investigate the \
phase-shift search with general but equal phase shifts. We are able to
considerably improve the algorithm by varying the phase-shift away from $\pi
/3$ when $\epsilon $ is large. As well known, the smaller deviation makes
the algorithm converge to the target state more rapidly. The deviation for
the Phase-$\pi /3$ search is $\epsilon ^{3}$\cite{Grover05}. For the large
size of database, we investigate that the deviation for any phase shifts of $%
\theta >\pi /3$ is smaller than $\epsilon ^{3}$ and the closer to $\pi $ the
phase shifts are, the smaller the deviation is. In this paper, we study the
performance of the general \ phase-shift search for only one iteration. This
also determines the failure probability and success probability of the
general \ phase-shift search after recursively applying the single iteration
for $n$ times. Note that we neglect the effects\ of decoherence completely
in this paper.

This paper is organized as follows. In section 3, we give the necessary and
sufficient conditions for the smaller deviation than $\epsilon ^{3}$. In
section 4, we show that the Phase-$\pi /3$ search algorithm performs well
for the small $\epsilon $. In section 6, we demonstrate that the closer to $%
\pi $ the phase shifts are, the smaller\ the deviation is. In section 7, we
propose the ratio measurement of the behavior of the Phase-$\theta $ search
algorithm for one query.

\section{Grover's \ phase-shift search and the reduction of the deviation}

The standard amplitude amplification algorithm would overshoot the target
state. To avoid drifting away from the target state, Grover presented the \
phase-shift search\cite{Grover05}.

In \cite{Grover05} the transformation $UR_{s}^{\pi /3}U^{+}R_{t}^{\pi /3}U$
was applied to the initial state $|s\rangle $,

\begin{eqnarray}
R_{s}^{\pi /3} &=&I-[1-e^{i\frac{\pi }{3}}]|s\rangle \langle s|,  \nonumber
\\
R_{t}^{\pi /3} &=&I-[1-e^{i\frac{\pi }{3}}]|t\rangle \langle t|,
\label{grover1}
\end{eqnarray}

\noindent where $|t\rangle $ stands for the target state. The transformation
$UR_{s}^{\pi /3}U^{+}R_{t}^{\pi /3}U$ is denoted as Grover's the Phase-$\pi
/3$ search algorithm in \cite{Tulsi}.

Grover let $\theta $ denote $\pi /3$. Then

\begin{eqnarray}
R_{s}^{\theta } &=&I-[1-e^{i\theta }]|s\rangle \langle s|,  \nonumber \\
R_{t}^{\theta } &=&I-[1-e^{i\theta }]|t\rangle \langle t|.  \label{grover2}
\end{eqnarray}

\noindent The transformation $UR_{s}^{\theta }U^{+}R_{t}^{\theta }U$ is
called as the Phase-$\theta $ search algorithm in this paper. As indicated
in \cite{Grover05}, when $\theta =\pi $, this becomes one iteration of the
amplitude amplification algorithm\cite{Grover98}\cite{Brassard97}. Note that
if we apply $U$ to the initial state $|s\rangle $, then the amplitude of
reaching the target state $|t\rangle $ is $U_{ts}$\cite{Grover98}.\ Applying
the transformation $UR_{s}^{\theta }U^{+}R_{t}^{\theta }U$ to the start
state $|s\rangle $, Grover derived the following,
\begin{equation}
UR_{s}^{\theta }U^{+}R_{t}^{\theta }U|s\rangle =U|s\rangle \lbrack
e^{i\theta }+\left\vert U_{ts}\right\vert ^{2}(e^{i\theta
}-1)^{2}]+|t\rangle U_{ts}(e^{i\theta }-1).  \label{grover3}
\end{equation}

Let $D(\theta )$ be the deviation from the $t$ state for any phase shifts of
$\theta $. Then from (\ref{grover3}) the following was obtained in \cite%
{Grover05},
\begin{equation}
D(\theta )=(1-\left\vert U_{ts}\right\vert ^{2})|e^{i\theta }+\left\vert
U_{ts}\right\vert ^{2}(e^{i\theta }-1)^{2}|^{2}.  \label{groverdev}
\end{equation}

\noindent Grover chose $\pi /3$ as phase shifts and let $\left\vert
U_{ts}\right\vert ^{2}=1-\epsilon $, where $0<\epsilon <1$. Substituting $%
\left\vert U_{ts}\right\vert ^{2}=1-\epsilon $, the deviation from the $t$
state becomes $D(\pi /3)=\epsilon ^{3}$\cite{Grover05}.

Deviation $D(\theta )$ in (\ref{groverdev}) can be reduced as follows. For
any $\theta $,
\begin{equation}
e^{i\theta }+\left\vert U_{ts}\right\vert ^{2}(e^{i\theta
}-1)^{2}=e^{i\theta }+2(\cos \theta -1)e^{i\theta }(1-\epsilon )=e^{i\theta
}[1+2(\cos \theta -1)(1-\epsilon )].  \label{reduction}
\end{equation}

\noindent So by (\ref{reduction}), we obtain
\begin{equation}
D(\theta )=\epsilon \lbrack 1+2(\cos \theta -1)(1-\epsilon )]^{2}.
\label{dev1}
\end{equation}

In this paper, we study the \ phase-shift search algorithm with two equal
phase shifts. It is clear that it is enough to consider $\theta $ in $[0,\pi
]$. It can be shown that the maximum and minimum of deviation $D(\theta )$
are $1$ and $0$. That is,
\begin{equation}
0\leq D(\theta )\leq 1.  \label{maxmin}
\end{equation}

\section{The phase shifts for smaller deviation}

As indicated in \cite{Grover98}, in the case of database search, $|U_{ts}|$
is almost $1/\sqrt{N}$, where $N$ is the size of the database. Thus, $%
\epsilon $ is almost $1-1/N$ and $\epsilon $ is close to $1$ for the large
size of database. It is known that the deviation for Grover's the Phase-$\pi
/3$ search is $\epsilon ^{3}$. In this section, we give the phase shifts for
smaller deviation than $\epsilon ^{3}$.

\subsection{Necessary and sufficient conditions}

From (\ref{dev1}) let us calculate
\begin{eqnarray}
D(\theta )-\epsilon ^{3} &=&\epsilon \lbrack 1+2(\cos \theta -1)(1-\epsilon
)]^{2}-\epsilon ^{3}  \nonumber \\
&=&\allowbreak \allowbreak \epsilon (1-\epsilon )(2\cos \theta -1)[2+(2\cos
\theta -3)(1-\epsilon )].  \label{dev2}
\end{eqnarray}

See Figs. 1 and 4. From (\ref{dev2}), we have the following statement.

Lemma 1. Deviation $D(\theta )$ in (\ref{dev1}) for any phase shifts of $%
\theta $ in $[0,\pi /3)$ is greater than $\epsilon ^{3}$ for any $\epsilon $%
. That is, $D(\theta )>\epsilon ^{3}$ for any $\theta $ in $[0,\pi /3)$ and
for any $\epsilon $. See table 1.

The argument is as follows.

When $0\leq \theta <\pi /3$, $0<2\cos \theta -1\leq 1$ and $2\epsilon
<2+(2\cos \theta -3)(1-\epsilon )\leq 1+\epsilon $ for any $\epsilon $.
Therefore when $0\leq \theta <\pi /3$, it follows (\ref{dev2})\ that $%
D(\theta )>\epsilon ^{3}$ for any $\epsilon $.

\ From (\ref{dev2}) and Lemma 1,\ the following lemma holds immediately. See
table 1.

Lemma 2. $D(\theta )<\epsilon ^{3}$ if and only if

\begin{eqnarray}
\theta >\pi /3\wedge \epsilon >1-2/(3-2\cos \theta ).  \label{cond2}
\end{eqnarray}

\noindent The following remark is used to describe the monotonicity of $%
1-2/(3-2\cos \theta )$ in (\ \ref{cond2}). The monotonicity is used to find
smaller deviation than $\epsilon ^{3}$ below.

Remark 1. $1-2/(3-2\cos \theta )$ increases from $-1$ to $3/5$ as $\theta $
increases from $0$ to $\pi $. Thus,

\begin{eqnarray}
-1\leq 1-2/(3-2\cos \theta )\leq 3/5.  \label{remark1}
\end{eqnarray}

Table 1. The phase shifts for deviations

\begin{tabular}{|c|c|c|}
\hline
$\theta $ &  & $\epsilon $ \\ \hline
When $\theta >\pi /3$ & $D(\theta )<\epsilon ^{3}$ & for $\epsilon
>1-2/(3-2\cos \theta )$ \\ \hline
When $\theta <\pi /3$ & $D(\theta )>\epsilon ^{3}$ & for any $\epsilon $ \\
\hline
When $\theta >\pi /3$ & $D(\theta )>\epsilon ^{3}$ & for $\epsilon
<1-2/(3-2\cos \theta )$ \\ \hline
\end{tabular}

\subsection{The phase shifts for smaller deviation}

In this subsection, we give the phase shifts for which the deviations are
smaller than $\epsilon ^{3}$. \

Corollary 1. Deviation $D(\theta )$ for any phase shifts of $\theta $ in $%
(\pi /3,\alpha ]$ is smaller than $\epsilon ^{3}$ whenever $\epsilon >$ $%
1-2/(3-2\cos \alpha )$.

Proof. By Remark 1, $1-2/(3-2\cos \theta )$ increases from $0$ to $%
1-2/(3-2\cos \alpha )$ as $\theta $ increases from $\pi /3$ to $\alpha $.
Thus, $0<1-2/(3-2\cos \theta )\leq $ $1-2/(3-2\cos \alpha )$\ whenever $\pi
/3<\theta \leq \alpha $. Therefore, when $\epsilon >$ $1-2/(3-2\cos \alpha )$%
, always $\epsilon $ $>1-2/(3-2\cos \theta )$. Hence, this corollary follows
Lemma 2.

When $\alpha =\pi $, $2\pi /3$, $\pi /2$\ and $\arccos \frac{1-3\delta }{%
2(1-\delta )}$, from Corollary 1 we have the following phase shifts for
smaller deviations than $\epsilon ^{3}$. See table 2.

Result 1. For any phase shifts of $\theta >\pi /3$, deviation $D(\theta
)<\epsilon ^{3}$ for $\epsilon >$ $3/5$. \ See Fig. 2 (a).

Result 2. For any phase shifts of $\theta $ in $(\pi /3,2\pi /3]$, deviation
$D(\theta )<\epsilon ^{3}$ for $\epsilon >1/2$. See Fig. 2 (b).

Result 3. For any phase shifts of $\theta $ in $(\pi /3,\pi /2]$, deviation $%
D(\theta )<\epsilon ^{3}$ for $\epsilon >1/3$. See Fig. 2 (c).

Result 4. When $\epsilon >$ $\delta $, for any phase shifts of $\theta $ in $%
(\pi /3$ ,$\arccos \frac{1-3\delta }{2(1-\delta )}]$, deviation $D(\theta
)<\epsilon ^{3}$.

Note that $\lim_{\delta \rightarrow 0}\arccos \frac{1-3\delta }{2(1-\delta )}%
=\pi /3$ .

Our conclusion is when we search large database, i.e., $\epsilon $ is large,
for any phase shifts of $\theta >\pi /3$ the deviation is smaller than $%
\epsilon ^{3}$.

Table 2. The phase shifts for $D(\theta )<\epsilon ^{3}$

\begin{tabular}{|c|c|c|}
\hline
$\theta $ &  & $\epsilon $ \\ \hline
When $\theta >\pi /3$ & $D(\theta )<\epsilon ^{3}$ & for $\epsilon >3/5$ \\
\hline
When $\pi /3<\theta \leq 2\pi /3$ & $D(\theta )<\epsilon ^{3}$ & for $%
\epsilon >1/2$ \\ \hline
When $\pi /3<\theta \leq \pi /2$ & $D(\theta )<\epsilon ^{3}$ & for $%
\epsilon >1/3$ \\ \hline
When $\pi /3<\theta \leq \arccos \frac{1-3\delta }{2(1-\delta )}$ & $%
D(\theta )<\epsilon ^{3}$ & for $\epsilon >\delta $ \\ \hline
\end{tabular}

\section{The Phase-$\protect\pi /3$ search is optimal for small
uncertainties.}

As indicated in \cite{Grover98}, the size of the database is very large,
i.e., $\epsilon $ is large. However, it is interesting to investigate the
performances of the Phase-$\pi /3$ search and the Phase-$\theta $ search for
small $\epsilon $.

\subsection{The Phase-$\protect\pi /3$ search possesses smaller failure
probability}

As said in \cite{Grover05}, $\epsilon ^{3}$ and $D(\theta )$ are the failure
probabilities of the Phase-$\pi /3$ search and the Phase-$\theta $ search,
respectively. Let us consider the ratio of the two failure probabilities. It
is easy to see that $\lim_{\epsilon \rightarrow 0}\epsilon ^{3}/D(\theta )=0$
for any $\theta \neq \pi /3$.\ That is, $\epsilon ^{3}=o(D(\theta ))$. In
other words, $\epsilon ^{3}$ is smaller than $D(\theta )$ for small $%
\epsilon $. It means that $\epsilon ^{3}$ approaches $0$ more rapidly than $%
D(\theta )$ as $\epsilon $ approaches $0$.

\subsection{The conditions under which the Phase-$\protect\pi /3$ search
behaves well \ \ \ \ \ \ \ \ \ \ }

Here, we discuss what $\epsilon $ satisfies $\epsilon ^{3}<$ $D(\theta )$.
From (\ref{dev2}) and Lemma 1, we have the following lemma.

Lemma 3. $D(\theta )>\epsilon ^{3}$ if and only if $\theta >\pi /3$ and $%
\epsilon <1-2/(3-2\cos \theta )$ or $0\leq \theta <\pi /3$. See table 1.

The following corollary follows Lemma 3.

Corollary 2. When $\pi /3<\alpha \leq \theta $ and $\epsilon <1-2/(3-2\cos
\alpha )$, $D(\theta )>\epsilon ^{3}$.

The argument is as follows. By Remark 1, $1-2/(3-2\cos \theta )$ increases
from $1-2/(3-2\cos \alpha )$ to $3/5$ as $\theta $ increases from $\alpha $
to $\pi $. Thus, when $\pi /3<\alpha \leq \theta $, $1-2/(3-2\cos \alpha
)\leq 1-2/(3-2\cos \theta )$. Consequently, this corollary holds by Lemma 3.

From Corollary 2 we have the following results.

Result 5. When $\theta \geq \pi /2$, $D(\theta )>\epsilon ^{3}$ for $%
\epsilon <1/3$.

Result 6. When $\theta \geq 2\pi /3$, $D(\theta )>\epsilon ^{3}$ for $%
\epsilon <1/2$.

Result 7. When $\theta \geq \arccos \frac{1-3\delta }{2(1-\delta )}$, $%
D(\theta )>\epsilon ^{3}$ for $\epsilon <\delta $.

Our conclusion is that for small $\epsilon $,\ the search algorithm performs
optimal for $\theta =\pi /3$. By means of the performance the Phase-$\pi /3$
search algorithm can be applied to quantum error corrections.

\section{Zero deviation and average zero deviation points}

\subsection{Zero deviation}

Let $d=1+2(\cos \theta -1)(1-\epsilon )$. Then, deviation $D(\theta )$ in (%
\ref{dev1})\ can be rewritten as $D(\theta )=\epsilon d^{2}$. Let $d=0$.
Then we obtain $\cos \theta =1-\frac{1}{2(1-\epsilon )}$, where $0<\epsilon
\leq \frac{3}{4}$ to make $\left\vert 1-\frac{1}{2(1-\epsilon )}\right\vert
\leq 1$. Conclusively, if $U_{ts}$ is given, that is, $\epsilon $ is fixed,
then we choose $\theta =\arccos [1-\frac{1}{2(1-\epsilon )}],$ which is in $%
(\pi /3$ ,$\pi ]$, as phase shifts. $\arccos [1-\frac{1}{2(1-\epsilon )}]$
will obviously make the deviation vanish and is called as a zero deviation
point. It means that one iteration will reach $t$ state with certainty if
the zero deviation point is chosen as phase shifts . Note that $%
\lim_{\epsilon \rightarrow 0}\arccos [1-\frac{1}{2(1-\epsilon )}]=\pi /3$ .
This says that $\pi /3$ is the limit of the zero deviation points $\theta $
though it is not a zero deviation point.

\subsection{Average zero deviation points}

When $0<\epsilon \leq \frac{3}{4}$, $\arccos [1-\frac{1}{2(1-\epsilon )}]$
is called as a zero deviation point. Since $\epsilon $ is not given, the
zero deviation point is unknown. However, if we know the range of $\epsilon $%
, then in terms of mean-value theorem for integrals, we can find the average
value $\bar{\theta}$ of the zero deviation points $\theta $. Here, we assume
that $\epsilon $ is uniformly distributed in the interval\ $(\beta ,\alpha
)\subseteq $ $(0,3/4]$.

Let $\epsilon $ be in the range $(\beta ,\alpha )$, where $(\beta ,\alpha
)\subseteq $ $(0,3/4]$. Then we calculate the average value of $1-\frac{1}{%
2(1-\epsilon )}$ over the range $(\beta ,\alpha )$ as follows.

\begin{eqnarray}
\frac{1}{\alpha -\beta }\int_{\beta }^{\alpha }[1-\frac{1}{2(1-\epsilon )}%
]d\epsilon =1+\frac{1}{2(\alpha -\beta )}\ln \frac{1-\alpha }{1-\beta }.
\end{eqnarray}

It can be argued that $-1\leq 1+\frac{1}{2(\alpha -\beta )}\ln \frac{%
1-\alpha }{1-\beta }<\frac{1}{2}$. Thus, it is reasonable to define

\begin{eqnarray}
\bar{\theta}=\arccos [1+\frac{1}{2(\alpha -\beta )}\ln \frac{1-\alpha }{%
1-\beta }].  \label{zeropoint}
\end{eqnarray}

\noindent $\bar{\theta}$ can be considered as the average value of the zero
deviation points $\theta $ and is called as the average zero deviation
point. It can be seen that $\pi /3<\bar{\theta}\leq \pi $.

When $\bar{\theta}$ is chosen as phase shift, we obtain the following
deviation
\begin{eqnarray}
D(\bar{\theta})=\epsilon (1+\frac{1-\epsilon }{\alpha -\beta }\ln \frac{%
1-\alpha }{1-\beta })^{2}. \label{zpdev}
\end{eqnarray}

Let us compute $D(\bar{\theta})-\epsilon ^{3}$ \ as follows.
\begin{eqnarray}
D(\bar{\theta})-\epsilon ^{3}=\epsilon (1+\frac{1}{\alpha -\beta }\ln \frac{%
1-\alpha }{1-\beta })(1-\epsilon )(1+\frac{1-\epsilon }{\alpha -\beta }\ln
\frac{1-\alpha }{1-\beta }+\epsilon ).
\end{eqnarray}

\noindent Notice that $1+\frac{1}{\alpha -\beta }\ln \frac{1-\alpha }{%
1-\beta }<0$ and $1-\frac{1}{\alpha -\beta }\ln \frac{1-\alpha }{1-\beta }>0$%
. Let $\kappa =1-2/(1-\frac{1}{\alpha -\beta }\ln \frac{1-\alpha }{1-\beta }%
) $. It can be proven that $0<\kappa <1$. We can conclude when $\epsilon
>\kappa $, $D(\bar{\theta})<\epsilon ^{3}$.

We will find the average zero deviation point $\bar{\theta}$ for the ranges $%
($ $0,1/2)$ and $(0,3/4)$ of $\epsilon $, respectively, as follows.

Example 1. Let $\epsilon $ lie in the range $(0,1/2]$.\ By (\ref{zeropoint}%
),\ the average zero deviation point $\bar{\theta}_{1}=\arccos (1-\ln
2)=72^{\circ }30^{\prime }$. Taking $\bar{\theta}_{1}$ as phase shifts, by (%
\ref{zpdev}) deviation $D(\bar{\theta}_{1})=\epsilon \lbrack 1-2(1-\epsilon
)\ln 2]^{2}$. Deviation $D(\bar{\theta}_{1})$ for phase shifts of $\bar{%
\theta}_{1}$ is smaller than $\epsilon ^{3}$, i.e., $D(\bar{\theta}%
_{1})<\epsilon ^{3}$, if and only if $\epsilon >\frac{2\ln 2-1}{2\ln 2+1}%
=\allowbreak 0.16$.

Example 2. Let $(0,3/4]$ be the range of $\epsilon $. Then by (\ref%
{zeropoint}), the average zero deviation point $\bar{\theta}_{2}$\ $=$ $%
\arccos (1-\frac{4}{3}\ln 2)=86^{\circ }$. Choosing $\bar{\theta}_{2}$ as
phase shifts,\ by (\ref{zpdev})\ deviation\ $D(\bar{\theta}_{2})={\small %
\epsilon \lbrack 1-}\frac{8}{3}{\small (1-\epsilon )}\ln {\small 2]}^{2}$
and $D(\bar{\theta}_{2})$\ is smaller than $\epsilon ^{3}$ when $\epsilon
>\allowbreak 0.30$. \

\section{Monotonicity of the deviation for large $\protect\epsilon $\ \ \ \
\ \ \ \ \ \ \ \ \ \ \ }

As discussed above, when $\epsilon $\ is fixed and lies in the range $%
(0,3/4] $ and $\arccos (1-\frac{1}{2(1-\epsilon )})$ is chosen as phase
shifts, the deviation vanishes. When $\epsilon >3/4$ ,\ since $\left\vert 1-%
\frac{1}{2(1-\epsilon )}\right\vert >1$, deviation $D(\theta )$\ does not
vanish for any phase shifts of $\theta $ in $[0,\pi ]$.

When $\epsilon \geq \frac{3}{4}$,

\begin{eqnarray}
-1\leq 2(\cos \theta -1)(1-\epsilon )\leq 0.  \label{cond1}
\end{eqnarray}

\noindent and $0\leq d\leq 1$. When $U_{ts}$ is given, that is, $\epsilon $
is fixed, by using (\ref{cond1}) it can be shown that deviation $D(\theta )$
monotonically decreases\ from $\epsilon $ to $\epsilon (4\epsilon -3)^{2}$
as $\theta $ increases from $0$ to $\pi $. See Fig. 3. for the monotonicity
of $D(\theta )$. When $\theta =\pi $, the deviation gets its minimum $%
\epsilon (4\epsilon -3)^{2}$. That is,
\begin{equation}
\epsilon (4\epsilon -3)^{2}\leq D(\theta )  \label{inequality}
\end{equation}

\noindent for any phase shifts of $\theta $ in $[0,\pi ]$, whenever $%
\epsilon \geq 3/4$.

Peculiarly,\textbf{\ }the deviation $\epsilon (4\epsilon -3)^{2}<$ $\epsilon
^{3}$ whenever $\epsilon >3/5$. The inequality in ({\ref{inequality})} also
follows that $(4\epsilon -3)\leq d$ for any phase shifts of $\theta $ in $%
[0,\pi ]$\ whenever $\epsilon \geq 3/4$.

See table 3 for the deviations $D(\theta )$\ for $\theta =\pi /2,2\pi
/3,3\pi /4,5\pi /6$,$\pi $. Also see Fig. 4.

Table 3. The deviations for $\epsilon >3/4$

\begin{tabular}{|c|c|c|c|c|}
\hline
${\small \theta }$ & ${\small \pi /2}$ & ${\small 2\pi /3}$ & ${\small 3\pi
/4}$ & ${\small 5\pi /6}$ \\ \hline
${\tiny D(\theta )}$ & ${\tiny \epsilon (2\epsilon -1)}^{2}$ & ${\tiny %
\epsilon (3\epsilon -2)}^{2}$ & ${\tiny \epsilon ((}\sqrt{2}{\tiny %
+2)\epsilon -(}\sqrt{2}{\tiny +1))}^{2}$ & ${\tiny \epsilon ((}\sqrt{3}%
{\tiny +2)\epsilon -(}\sqrt{3}{\tiny +1))}^{2}$ \\ \hline
\end{tabular}

Remark 2.

From the discussion above, it is easy to see that the closer to $\pi $ the
phase shifts are, the smaller\ the deviation is, when $\epsilon \geq \frac{3%
}{4}$. By means of the inequality in ({\ref{inequality}) we can discuss the
lower bound of the number of iterations to find the $t$ state. }

Note that when the selective phase shift $\theta $ becomes $\pi $, the phase-%
$\pi $ search is the amplitude amplification search.

\section{The ratio measurement of the success probabilities for one query}

\subsection{The ratio of the success probabilities}

Clearly, the greater the success probability is, the better the algorithm
performs. In other words, the more rapidly the algorithm converges. In this
section, it is demonstrated that the limit of the ratio of success
probabilities of the Phase-$\theta $ and the Phase-$\pi /3$ search
algorithms is used to quantify the performance of the Phase-$\theta $ search
algorithm.

From (\ref{maxmin}), let $\Delta (\theta )$ $=$ $1-D(\theta )$. Then $\Delta
(\theta )$ is the success probability with which the transformation $%
UR_{s}^{\theta }U^{+}R_{t}^{\theta }U$ in (\ref{grover3}) drives the start
state to the target state. For instance, $\Delta (\pi /3)=1-D(\pi
/3)=1-\epsilon ^{3}$, which is the success probability of the Phase-$\pi /3$
search algorithm for one query. See Page 1 in \cite{Grover05}. Explicitly, $%
\Delta (\theta )$ is not the desired measurement free of $\epsilon $ for the
Phase-$\theta $ search algorithm because $\Delta (\theta )$ is also a
function of $\epsilon $.

Let us compute the limit of $\Delta (\theta )$ as $\epsilon $ approaches 1
as follows.

$\lim_{\epsilon \rightarrow 1}\Delta (\theta )=\lim_{\epsilon \rightarrow
1}(1-\epsilon (1+2(\cos \theta -1)(1-\epsilon ))^{2}))=0$, for any $\theta $
in $[0,\pi ]$.

It is straightforward that the above limit can not be used to describe the
performance of the Phase-$\theta $ search algorithm for any phase shifts of $%
\theta $ in $[0,\pi ]$\ because the limit always is zero for any $\theta $
in $[0,\pi ]$.

It is natural to consider and calculate $\frac{\Delta (\theta )}{\Delta (\pi
/3)}$ as follows.

\begin{eqnarray}
\frac{\Delta (\theta )}{\Delta (\pi /3)}=\frac{4\left( \cos ^{2}\theta
\right) \epsilon ^{2}-8\left( \cos \theta \right) \epsilon ^{2}+4\epsilon
^{2}+4\left( \cos \theta \right) \epsilon -4\left( \cos ^{2}\theta \right)
\epsilon +1}{\epsilon ^{2}+\epsilon +1}.
\end{eqnarray}

\noindent Then we obtain the following limit of $\frac{\Delta (\theta )}{%
\Delta (\pi /3)}$ as $\epsilon $ approaches 1. Let

\begin{eqnarray}
\rho =\lim_{\epsilon \rightarrow 1}\frac{\Delta (\theta )}{\Delta (\pi /3)}%
=\allowbreak \frac{5-4\cos \theta }{3}.  \label{rate}
\end{eqnarray}

\noindent Then $\rho $ can be considered as the ratio of success
probabilities for the Phase-$\theta $ and the Phase-$\pi /3$ search
algorithms for large $\epsilon $.\ Notice that $\rho $ is free of $\epsilon $
and only depends on $\theta $. Hence, $\rho $ can be considered as a
measurement of performance of Phase-$\theta $ search algorithm for any phase
shifts of $\theta $\ in $[0,\pi ]$.

We can follow \cite{Grover05} to define by the recursion $U_{m+1}=$ $%
U_{m}R_{s}^{\theta }U_{m}^{+}R_{t}^{\theta }U_{m}$, where $U_{0}=U$. For the
Phase-$\pi /3$ search, after recursive application of the basic iteration $m$
times, the success probability $\left\vert U_{m,ts}\right\vert =1-\epsilon
^{3^{m}}$\cite{Grover05}. For the Phase-$\theta $ search, as well we can
derive the success probability $\left\vert U_{m,ts}\right\vert $ and the
failure probability $1-\left\vert U_{m,ts}\right\vert $ after recursive
application of the basic iteration $m$ times. Fixed points of the Phase-$%
\theta $ search algorithm are discussed in \cite{reviewer}.

\subsection{The larger phase shifts than $\protect\pi /3$ for larger size of
database}

It can be shown that $\rho $ increases from $1/3$ to $3$ as $\theta $
increases from 0 to $\pi $. In particular, $\rho $ increases from $1$ to $3$
as $\theta $ increases from $\pi /3$ to $\pi $. This also says that for
large databases, the larger the phase shifts are, the greater the success
probabilities are. For instance, $\rho =2.8$ for Phase-$5\pi /6$ search.
This means that for large $\epsilon $, the ratio of success probabilities
for the Phase-$5\pi /6$ and the Phase-$\pi /3$ search is 2.8. See table 4.

Table 4.\ $\rho $'s values for the Phase-$\theta $ search

\begin{tabular}{|c|c|c|c|c|c|}
\hline
$\theta $ & $\pi /2$ & $2\pi /3$ & $3\pi /4$ & $5\pi /6$ & $\pi $ \\ \hline
$\rho $ & $5/3$ & $7/3$ & $(5+2\sqrt{2})/3=2.6$ & $(5+2\sqrt{3})/3=2.8$ & $3$
\\ \hline
\end{tabular}%
\ \ \ \ \ \ \ \ \ \ \

\section{Summary}

In this paper, we give the phase shifts for smaller deviation than $\epsilon
^{3}$. When $\epsilon \leq 3/4$ and $\epsilon $ is given, we choose the zero
deviation point as phase shifts to find the desired state for one iteration.
When $\epsilon \geq 3/4$, the deviation decreases from $\epsilon ^{3}$\ to $%
\epsilon (4\epsilon -3)^{2}$ as $\theta $ increases from $\pi /3$ to $\pi $.
It is shown that for small $\epsilon $, the Phase-$\pi /3$ search behaves
better than the general Phase-$\theta $ search. Therefore the Phase-$\pi /3$
search can be applied to quantum error correction. We propose the limit of
the ratio of success probabilities of the Phase-$\theta $ and the Phase-$\pi
/3$ search algorithms as a measure of efficiency of a single Phase-$\theta $
iteration. The measure can help us find the optimal phase shifts for small
deviation and large success probability. Thus, there are more choices for
phase shifts to adjust an algorithm for large size of database and more
loose constraint opens a door for more feasible or robust realization.

Acknowledgement

We want to thank Lov K. Grover for his helpful discussions and
comments on the original manuscript ( in December, 2005) and the
reviewer for the helpful comments on this paper and useful
discussions about fixed points of the Phase-$\theta $ search.


\begin{thebibliography}{99}
\bibitem{reviewer} In the reviewer's report on this paper, the reviewer
investigated fixed points of the Phase-$\theta $ search algorithm as
follows. The reviewer let $f(\epsilon _{n})=\epsilon _{n+1}=\epsilon
_{n}[1+2(cos\theta -1)(1-\epsilon _{n})]^{2}$. Then the reviewer showed that
the equation has $0$ to be an attractive fixed point when $0<\theta <\pi /2$
because $0<f^{\prime }(0)<1$, $\cos \theta /(\cos \theta -1)$ to be an
attractive fixed point when $\pi /2<\theta <2\pi /3$ because $|f^{\prime
}(\cos \theta /(\cos \theta -1))|<1$, and all the fixed points to be
repulsive when $2\pi /3<\theta <\pi $. Following the reviewer's idea, we can
show when $\theta =\pi /2$, $0$ is a semi-attractive fixed point because $%
f^{\prime }(0)=1$ and $f^{\prime \prime }(0)=-8$ and when $\theta =2\pi /3$,
$1/3$ is also a semi-attractive fixed-point because $f^{\prime }(1/3)=-1$
and $f^{\prime \prime }(1/3)=-6$.
\end{thebibliography}
\end{document}